# Synchronization of spin-transfer torque oscillators by spin pumping, inverse spin Hall, and spin Hall effects

Mehrdad Elyasi, Charanjit S. Bhatia, and Hyunsoo Yang[a]

*Department of Electrical and Computer Engineering, National University of Singapore, 4 Engineering Drive 3, Singapore 117576, Singapore*

We have proposed a method to synchronize multiple spin-transfer torque oscillators based on spin pumping, inverse spin Hall, and spin Hall effects. The proposed oscillator system consists of a series of nano-magnets in junction with a normal metal with high spin-orbit coupling, and an accumulative feedback loop. We conduct simulations to demonstrate the effect of modulated charge currents in the normal metal due to spin pumping from each nano-magnet. We show that the interplay between the spin Hall effect and inverse spin Hall effect results in synchronization of the nano-magnets.

[a] E-mail: eleyang@nus.edu.sg

## I. INTRODUCTION

Magnetic nano-oscillators have recently emerged from the spin-transfer torque (STT) phenomenon.[1-4] Nano-contact or patterned spin-valves and magnetic tunnel junctions (MTJ) have been pursued for achieving spin torque oscillators (STO).[5-8] Since the power generated by a single nano-oscillator is not able to satisfy the requirements of its targeted applications, there have been many efforts in order to achieve a synchronization method of multiple oscillators.[5,9-14] In order to synchronize several STOs, a feedback system that can enforce the oscillations to lock to achieve the minimum free energy should exist. Kaka *et al.* demonstrated the first synchronization between two nano-STOs based on spin waves interactions,[5] in which the feedback system was mediated by spin waves. Injection locking is another approach in which an external alternating current is used to lock a group of STOs.[9,14-16] The injection locking has been inspiration to pursue synchronization in MTJs or spin-valve STOs in series connection, where the stimulated current due to the oscillating magnetoresistance is utilized to form the feedback system and induce locking.[11,17] Difficulties of fabricating spin-valve or MTJ STOs in series connection have been an obstacle to demonstrate the performance of this synchronization method. Maximum four nano-contact STOs have been experimentally synchronized so far mediated by anti-vortices.[13] However, a relatively low frequency of vortex oscillators and high current required to induce vortex oscillations are disadvantages of this method.

The spin Hall effect (SHE) has enabled generation of pure spin currents from charge currents in a high spin orbit coupled material.[18] A charge current is induced in a normal metal due to the inverse spin Hall effect (ISHE), when a spin current is injected into the normal metal.[19-21] The conversion efficiency (spin Hall angle) between a charge current and spin current in the SHE and ISHE processes depends on the strength of spin-orbit scattering in the normal metal. Spin transfer ferromagnetic resonance (ST-FMR), spin pumping, and time-resolved magneto-optical Kerr effect (TR-MOKE) experiments have been pursued in order to quantify the spin Hall angle of different heavy metals such as Pt and Ta.[20,22-26] Pure spin currents generated by the SHE have been implemented to switch a magnetic layer.[27-30] Moreover, a new type of oscillator based on the SHE has been introduced, in which the oscillation mode is a self-localized spin wave bullet,[31-33] and a recent experiment has demonstrated its injection locking.[34] The spin Hall oscillators (SHO) can be also realized in nano-magnetic elements as an auto oscillation mode.[35]

A method to synchronize multiple STOs without requiring an external ac excitation is still a subject of study. In this work, we propose a synchronization method based on the spin pumping, SHE, and ISHE. The device consists of nano-magnets in junction with a normal metal with high spin-orbit coupling. We implement an accumulative feedback loop for the modulated part of the current. Numerical simulations have been carried out to study the effect of the modulated charge current due to the spin pumping and ISHE, on the magnetization oscillation of nano-magnets. It is found that under certain circumstances, synchronization of the oscillation of multiple nano-magnets is possible.

## II. MODELING METHOD

The proposed oscillator system consists of $N$ elliptic ferromagnetic single domain elements in junction with a high spin-orbit scattering non-magnetic (NM) metal shown in Fig. 1(a). Magnetic elements ($F_i$ to $F_{i+1}$) are assumed to be magnetically isolated. The NM strip is as wide as the long axis of the magnetic elements. If a charge current ($\vec{J}_{in}$) is applied to the device, it gives rise to a transverse pure spin current due to the SHE (see Fig. 1(d)). The generated spin current induces magnetization oscillations in the magnetic elements. Figures 1(b) and 1(c) demonstrate the configuration of the initial magnetization in the $i^{th}$ magnetic element ($\vec{M}_{0_i}$) and the spin polarization direction ($\hat{s}$) of the pure spin current generated by the SHE. The magnetization precession in each magnetic element pumps spins into the NM metal. Subsequently, the pumped pure spin current is converted back to a charge current due to the ISHE (see Fig. 1(e)). Therefore, the initial charge current is modulated with the ISHE-generated charge current. The modulated charge current again generates a pure spin current due to the SHE. Therefore, a feedback is formed from each magnet to all others through a sequence of the spin pumping, ISHE, and SHE. By using the conventional feedback loop,[11,12,17,36] the modulated spin current is not strong enough to cause any locking of the oscillations due to the small spin-Hall angle ($\theta_{SH}$). To increase the locking strength, there is a need for amplification or accumulation of the modulated part of the charge current. Figure 1(f) demonstrates such a feedback loop implemented in the oscillator system. The components of the feedback are two bias tees that mix or decouple the dc and ac currents, a line which has a specific time delay $\tau$, an amplifier with a gain $G$ and maximum output current density of $J_M$, and a switch which is activated at a certain time ($\tau_{switch}$) after the magnetization oscillation is initiated by the dc current. It is worthy to note that similar feedback loops were utilized for generation of spin wave solitons.[37,38]

The Landau-Lifshitz-Gilbert (LLG) equation including the Slonczewski spin-transfer term,[1] which governs the magnetization dynamics of each magnetic element is

$$\frac{d\vec{m}_i}{dt} = -\gamma \vec{m}_i \times \vec{H}_{eff_i} + \alpha \vec{m}_i \times \frac{d\vec{m}_i}{dt} + \frac{\gamma \hbar \varepsilon}{\mu_0 e d M_s} J_{s_i} (\vec{m}_i \times \hat{s} \times \vec{m}_i) \quad (1)$$

where $\vec{m}_i$ is the normalized magnetization vector ($\vec{M}_i/M_s$) of the $i^{th}$ element, $\gamma$ is the gyromagnetic ratio, $\vec{H}_{eff_i}$ is the effective field, $\alpha$ is the Gilbert damping constant, $\hbar$ is the reduced plank constant, $\varepsilon$ is the efficiency of the spin transfer, $\mu_0$ is the permeability of vacuum, $e$ is the electron charge, $d$ is the thickness of the magnetic element, $M_s$ is the saturation magnetization, and $J_{s_i}$ is the spin current injected into the $i^{th}$ magnetic element and is polarized in the direction of $\hat{s}$. The effective field for the $i^{th}$ element, $\vec{H}_{eff_i}$ can be written as $\vec{H}_{eff_i} = \vec{H}_{d_i} + \vec{H}_{ext} + \vec{H}_{an_i}$, where $\vec{H}_{d_i}$ is the demagnetizing field equal to $-M_s(\vec{m}_i \cdot \hat{z})\hat{z}$, $\vec{H}_{ext}$ is the external field in the $x$ direction, and $\vec{H}_{an_i}$ is the shape anisotropy field defined as $H_{an_i}(\vec{m}_i \cdot \hat{x})\hat{x}$.

The spin current ($\vec{J}_{sp_i}$) which is pumped into the NM metal from magnetization dynamics of the $i^{th}$ magnetic element can be defined as $\vec{J}_{sp_i} = \frac{\hbar}{8\pi} Re(2 g_{\uparrow\downarrow}) \left[ \vec{m}_i \times \frac{d\vec{m}_i}{dt} \right]$, where $g_{\uparrow\downarrow}$ is the spin mixing conductance in the interface of the ferromagnetic material and the NM metal. The charge current generated in the NM metal from $\vec{J}_{sp_i}$ due to the ISHE is defined as $\vec{J}_{c_i} = -\frac{e}{\hbar} \theta_{SH} \left[ \vec{J}_{sp_i} \times \vec{m}_i \right]$. The input current to the oscillator system, $\vec{J}_{in}$ at time $t$, can be defined as $\vec{J}_{in}(t) = -J_{dc}\hat{y} + \sum_{i=1}^{N} \sum_{n=1,2,...} \left( \vec{J}_{c_i}(t-n\tau) \cdot \hat{y} \right) \times G^n (1-L_{line})^n$, where $J_{dc}$ is the amplitude of the dc charge current, and $L_{line}$ is the line loss. Similarly, the output current of the oscillator system is $J_{out}(t) = \vec{J}_{in}(t) + \sum_{i=1}^{N} \vec{J}_{c_i}(t) \cdot \hat{y}$. The amplifier sets a limit for its maximum output current as $J_M$. Therefore, if the ac part of $\vec{J}_{out}$ satisfies $|\vec{J}_{ac,out}| > J_M$, $\vec{J}_{in}(t) = -J_{dc}\hat{y} + J_M \times \frac{\vec{J}_{ac,out}(t)}{|\vec{J}_{ac,out}(t)|}$. The spin current ($J_{s_i}$) which is injected in each element and its polarization ($\hat{s}$) are derived considering the SHE of the charge current $J_{out}$ as $J_{s_i}\hat{s} = \theta_{SH}\left[\vec{J}_{out}(t) \times \hat{z}\right]$.

We assume a magnetic material with a saturation magnetization ($M_s$) of $7 \times 10^5$ A/m, Gilbert damping ($\alpha$) of 0.007 (for permalloy[22]), and gyromagnetic ratio $\gamma/2\pi$ of $3.5 \times 10^4$ Hz·(m/A). The spin Hall angle ($\theta_{SH}$) of the NM metal is assumed to be 0.07 (for Pt[22]). The spin mixing conductance ($g_{\uparrow\downarrow}$) is assumed to be $2.1 \times 10^{19}$/m$^2$ as previously reported for the Py/Pt interface.[20] The thickness ($d$) of the magnetic layer is chosen as 5 nm. The initial magnetization of the magnetic elements are assumed to be $\theta_i = 75°$ and $\varphi_i = 5°$. To calculate the magnetization dynamics of each magnetic element, we solve Eq. (1) using Runge-Kutta numerical method.[39]

## III. RESULTS AND DISCUSSION

Figure 2 demonstrates an example of synchronization in frequency and phase using the proposed method. For this example, the number of the magnetic elements (N) is 10, $\tau$ is 5 ps, $\tau_{switch}$ is 10 ns, $G$ is $1/(1-L_{line})$, $L_{line}$ is 0.1, the external magnetic field $H_{ext}$ is 500 Oe, and the dc current $J_{dc}$ is $(4 \times 10^{10})/\theta_{SH}$ A/m$^2$. The efficiency of the spin-transfer ($\varepsilon$) is assumed to be unity. To introduce different oscillation frequencies among the elements, shape anisotropy $H_{an_i}$ is defined as $H_{an_i} = h_{an} + \left[h_{dev} \times (i-1)\right]/(N-1)$, where $h_{an}$ and $h_{dev}$ are the minimum value and the maximum deviation of shape anisotropy fields, respectively. For the case in Fig. 2, $h_{an}$ is 50 Oe and $h_{dev}$ is 15 Oe. Figure 2(a) shows the temporal variation of $\vec{m}_T$ and $\vec{m}_{T_{sync}}$ that are summations of the magnetization of all the elements ($\sum_{i=1}^{N} \vec{m}_i$), without (black curve) and with (red curve) applying ISHE-SHE interaction among the elements and the described feedback, respectively. The line-like collective trajectory of the red curve is the signature of synchronization. In addition, synchronization is clearly shown in Fig. 2(b) which depicts the fast Fourier transforms (FFT), $\left\| \sum_{\hat{e}_i = \hat{x},\hat{y},\hat{z}} FFT(\vec{m}_T \cdot \hat{e}_i) \right\|$ and $\left\| \sum_{\hat{e}_i = \hat{x},\hat{y},\hat{z}} FFT(\vec{m}_{T_{sync}} \cdot \hat{e}_i) \right\|$. A significant reduction of the linewidth (with the factor of ~1/N where N = 10) and increase in the spectral density peak (with the factor of ~N) are achieved due to synchronization. Note that the power spectral density (PSD) peak scales with ~$N^2$ and its linewidth scales with ~$1/N^2$. Figure 2(c) shows the change in the ac part of $J_{out}$ ($J_{ac,out}$ in Fig. 1(f)) for the synchronized case. Stabilizing of $J_{ac,out}$ is the consequence of the frequency and phase locking of the elements after 15 ns.

The switching delay of the ac feedback line, $\tau_{switch}$ can affect the synchronization, since it determines the stage of the oscillation, in which the magnets begin to interact.[40] In all the modeling results, we include the feedback system after 10 ns ($\tau_{switch}$ = 10 ns), which is well after all the oscillators have achieved their individual equilibrium precession. For in-plane magnetized films and in-plane polarized spin currents, in-plane (IP) and out-of-plane (OOP) precessions can be achieved.[41] Since the locking bandwidth for IP precession is very low in comparison with the OOP precession,[11,12] we confine our analysis to the OOP oscillation. For the fields used in Fig. 2, our calculations show that the OOP mode emerges for $J_{dc} > (3.9 \times 10^{10})/\theta_{SH}$ A/m$^2$.

The effect of the time delay ($\tau$) of the feedback line has been an important attribute of the injection locking, as the magnetic oscillators are locked to their own oscillation with a time delay.[11,12] The injection of an ac current into a STO enforces its oscillation to have a phase difference of ~ $\pi/2$ with respect to the injection.[15] Therefore, in the conventional feedback loop,[11,12,17,36] it is expected to achieve locking when $\tau$ is in the range of a quarter of the oscillation period of the magnets. For the case of the accumulative feedback loop (see Fig. 1(f)), the relation between the feedback signal phase and $\tau$ is different. As an example, we assume an oscillatory function containing three harmonics with typical amplitude ratios, $A(t) = \sin(\omega t) + 0.3\sin(2\omega t) + 0.1\sin(3\omega t)$, with $\omega = 2\pi f$ where $f = 6$ GHz. Figure 3(c) shows the variation of the difference of the phase ($\varphi_{diff}$) between the feedback signal $B(t) = \sum_{n=1}^{1000} A(t - n\tau)$ and the original signal A(t) with respect to the variation of $\tau$ (red curve). The $\varphi_{diff}$ for the accumulative delay is close to $\pi/2$ for values of $\tau = \sim 0$, $q/f$ (q = 1, 2, …), whereas for a single feedback (blue curve, $B(t) = A(t - \tau)$) $\varphi_{diff}$ is close to $\pi/2$ for values of $\tau = \sim [q/f + 1/(4f)]$ (q=0, 1, …). Figures 3(a) and (b) demonstrate the synchronization behavior with a variation of $\tau$, where $N = 3$, $J_{dc} = (5 \times 10^{10})/\theta_{SH}$ A/m$^2$, and $h_{dev} = 10$ Oe. It can be seen that the synchronization is realized for $1 < \tau < 12$ ps ($\tau \approx 0$) and $156 < \tau < 168$ ps ($\tau \approx 1/f$), where $1/f \approx 159$ ps, in line with the above simple analytical model.

In Fig. 3(d), we calculate the nonlinearity coefficient $\nu = \dfrac{d\omega/dp}{d\Gamma_+/dp - d\Gamma_-/dp}$ for the dc currents corresponding to the OOP oscillation, where $p$ is the oscillation power, $\Gamma_+$ is the damping torque, and $\Gamma_-$ is the anti-damping torque.[42] For the currents leading to OOP precession $|\nu|$ is high, therefore, the internal phase shift is $\varphi_{int} = -\tan^{-1}(\nu) \approx \pi/2$. The shift in the frequency for the locked cases in Fig. 3(a) can be described utilizing the simple non-linear oscillator model, which results in $\omega_{sync} = \omega_0 + \Omega \sin(\varphi_{diff} - \varphi_{int})$, where $\omega_{sync}$ is the synchronized frequency, $\omega_0$ is the free running frequency of the oscillator, and $\Omega$ is the feedback strength.[17,42]

Relatively small values of $J_M$ (maximum current density in the feedback line) may clip the input signal to the feedback loop ($J_{ac,out}$), and consequently there will be additional frequency components in the output of the feedback ($J_{ac,in}$). Therefore, sidebands of the main harmonics will appear in the oscillations. For example, additional frequency components appear in Fig. 2(b) in the case of $J_M = 1.5 \times 10^{10}$ A/m$^2$ (blue curve) due to a relatively small value of $J_M$. As $N$ is higher, a larger $J_M$ is required for the synchronization to be realized.

In order to study the relation of the dc current ($J_{dc}$) and the locking bandwidth (BW), we have varied $h_{dev}$ from 4 to 40 Oe for different currents leading to OOP precession with N = 3 and $\tau$ = 5 ps. Figure 4(a) shows the first harmonic of $\left\| \sum_{\hat{e}_i = \hat{x}, \hat{y}, \hat{z}} FFT \left( \left( \sum_{i=1}^{N} \vec{m}_i \right) \cdot \hat{e}_i \right) \right\|$ for various $h_{dev}$ and $J_{dc}$ combinations. It has been described in theoretical studies that for small oscillation powers, the maximum locking bandwidth is $\Omega\sqrt{1+\nu^2}$.[42] From the values of $\nu$ presented in Fig. 3(d), and relating it to the locking BW variation with respect to the current, a peak in the locking BW at $J_{dc} \approx 4.5 \times 10^{10}/\theta_{SH}$ A/m$^2$ is expected. However, from a complex behavior of locking BW in Fig. 4(a), it can be inferred that the feedback strength $\Omega$ has a fluctuating variation with respect to $J_{dc}$. The fluctuation is due to the variation of mean oscillation angle and the oscillation trajectory with respect to dc current, which through spin pumping, ISHE, the accumulative feedback line, and SHE cause fluctuation in the feedback strength.

Figure 4(b) shows the locking behavior with a higher $\theta_{SH} = 0.3$ reported for β-W[43] at $J_{dc}$ = $5\times10^{10}/\theta_{SH}$ A/m$^2$, assuming the same intermixing conductance ($g_{\uparrow\downarrow}$) as the Pt/Py bilayer for the β-W/Py bilayer. By comparing Fig. 4(b) with the corresponding $J_{dc}$ case in Fig, 4(a), it is found that the locking BW as well as the locked frequency are increased due to a higher amplitude of the feedback signal which is proportional to $\theta_{SH}^2$. Note that as the feedback signal increases, a higher $J_M$ is required.

Finite temperature will induce a random field with Gaussian distribution in both time and space with a strength of $\sigma^2 = 2\alpha kT / \gamma M_s V$, where $k$ is the Boltzmann constant, $T$ is the temperature, and $V$ is the volume of the magnetic element.[44] Figure 4(c) shows the locking behavior with considering such a random field for different values of $\sigma_T = \sigma / \mu_0$. For elements with exemplary dimensions of 100 nm × 40 nm × 5 nm at $T$ = 300 K, $\sigma_T$ = 3.08 Oe. The Joule heating will increase the temperature $T$, resulting in a higher thermal noise, however, Fig. 4(c) shows that locking can be achieved for high thermal noise values ($\sigma_T$ = 5 Oe corresponds to $T$ = 786 K for the exemplary dimensions).

The fabrication of the proposed system requires patterning of a narrow strip of a NM metal such as Pt or Ta, and ferromagnetic nano-magnets such as NiFe or CoFeB. For an effective modulation of the current in NM and subsequent effective synchronization, the width of the magnetic elements should be the same as the width of the strip, and the thickness of the NM should be less than its spin diffusion length. This fabrication process is much simpler than that of MTJ based STOs in series connection which has not been demonstrated experimentally. Moreover, the proposed method can be realized in a 3-terminal configuration[35] by substituting the nano-magnets with nano-patterned MTJs with their free layers adjacent to the NM metal. In addition, the feedback loop implemented in this work can be utilized for the synchronization of spin-valve based STOs which generate a low charge current modulation.

In summary, we have proposed a method to synchronize multiple nano-magnets in the frequency and phase using the spin pumping, inverse spin Hall, and spin Hall effects. By adding an ac feedback line which imposes a certain delay time and gain, synchronization is realized for a wide range of applied dc currents. The synchronization is achieved for the feedback delay times close to 0 or integer multiples of the oscillation period. It offers a high power phase locked oscillation system with a narrow linewidth and its frequency can be controlled in a wide range. Our results intrigue future experimental studies for the demonstration of the proposed system.

**ACKNOWLEDGMENTS**

This research is supported by the National Research Foundation, Prime Minister's Office, Singapore under its Competitive Research Programme (CRP Award No. NRF-CRP 4-2008-06 and NRF-CRP12-2013-01).

**Fig. 1.** (a) The schematic diagram of the oscillator system. (b) The configuration of the initial magnetization, spin current polarization, and the external field for the $i^{th}$ element. (c) The polar ($\theta_i$) and azimuthal ($\varphi_i$) angles of the initial magnetization of the $i^{th}$ element. (d) The schematic representation of the SHE. (e) The schematic diagram of the spin pumping and ISHE. $\vec{m}_i$, $\vec{J}_{c_i}$, and $\vec{J}_{sp_i}$ form a right handed coordinate. (f) The schematic of the complete oscillator system including the ac feedback line.

**Fig. 2.** (a) The temporal variation of $\vec{m}_T$ (black curve) and $\vec{m}_{T_{sync}}$ (red curve). $m_{x,(y,z)}$ is the $x,(y,z)$ component of $\vec{m}_T$ or $\vec{m}_{T_{sync}}$. (b) The spectral density (SD) of FFT of $\vec{m}_T$ and $\vec{m}_{T_{sync}}$. (c) The time variation of the ac part of $J_{out}$. $H_{ext}$ = 500 Oe, $h_{an}$ = 50 Oe, $h_{dev}$ = 15 Oe, $N$ = 10, $J_{dc}$ = $(4 \times 10^{10})/\theta_{SH}$ A/m$^2$, and $\tau$ = 5 ps are used. $J_M$ = $3 \times 10^{10}$ A/m$^2$ for the red curves in (a) and (b) and the curve in (c), while $J_M$ = $1.5 \times 10^{10}$ A/m$^2$ for the blue curve in (b).

**Fig. 3.** (a) Effect of $\tau$ variation on synchronization for $H_{ext}$ = 500 Oe, $h_{an}$ = 50 Oe, $h_{dev}$ = 10 Oe, $J_{dc}$ = $(5 \times 10^{10})/\theta_{SH}$ A/m$^2$, $N$ = 3, and $J_M$ = $1 \times 10^{10}$ A/m$^2$. (b) Spectral density (SD) versus $\tau$. (c) $\varphi_{diff}$ versus $\tau$ for the normal feedback (blue curve) and the accumulative (Acc.) feedback (red curve). (d) Nonlinearity coefficient $\nu$ versus $J_{dc}$.

**Fig. 4.** (a) Effect of the dc current ($J_{dc}$) on the maximum possible $h_{dev}$ for synchronization (locking bandwidth (BW)). $H_{ext}$ = 500 Oe, $h_{an}$ = 50 Oe, $N$ = 3, $\tau$ = 5 ps, and $J_M$ = $1 \times 10^{10}$ A/m$^2$ are used. (b) Locking dependence on $h_{dev}$ for $\theta_{SH}$ = 0.3 with $J_{dc}$ = $5 \times 10^{10}/\theta_{SH}$ A/m$^2$ and $J_M$ =

$3 \times 10^{10}$ A/m$^2$. (c) Locking possibility for different $\sigma_T$ with $\theta_{SH} = 0.07$, $J_{dc} = 5 \times 10^{10}/\theta_{SH}$ A/m$^2$, and $J_M = 1 \times 10^{10}$ A/m$^2$. SD is the normalized spectral density of the FFT of $\vec{m}_{T_{sync}}$.

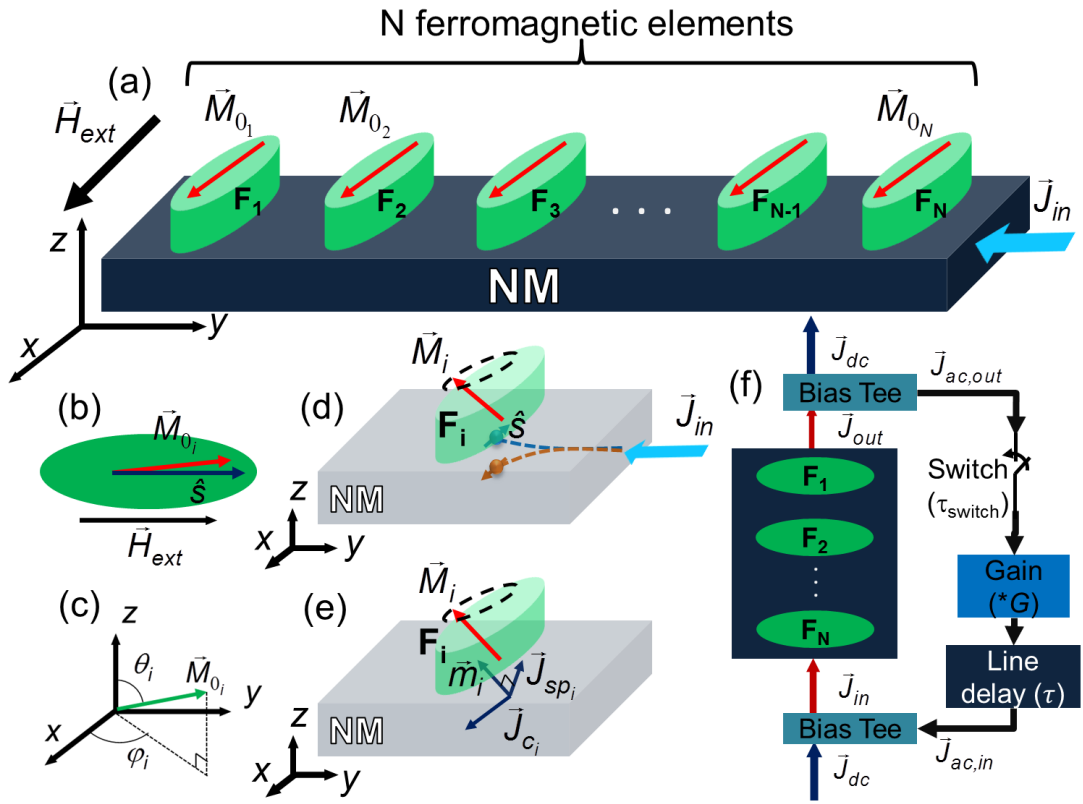

Fig. 1

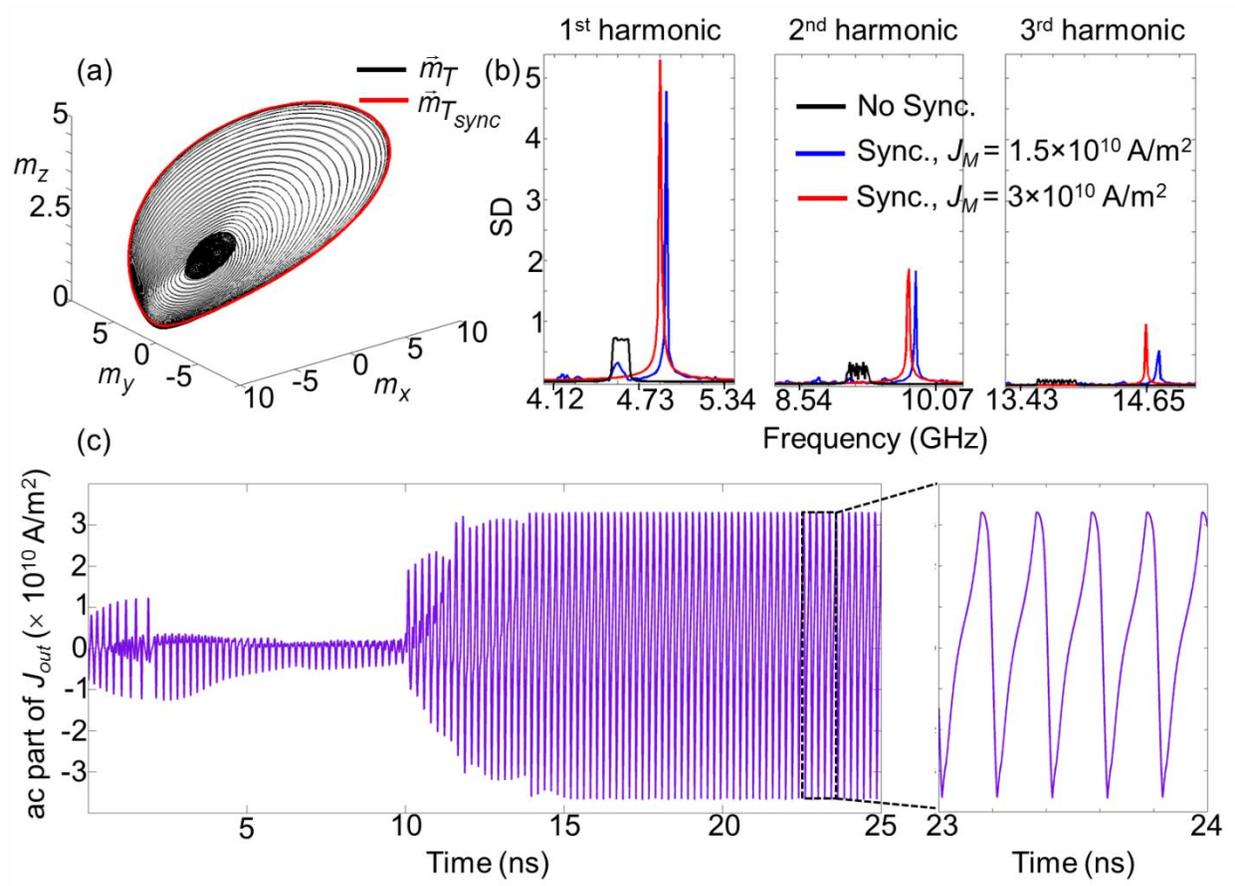

Fig. 2

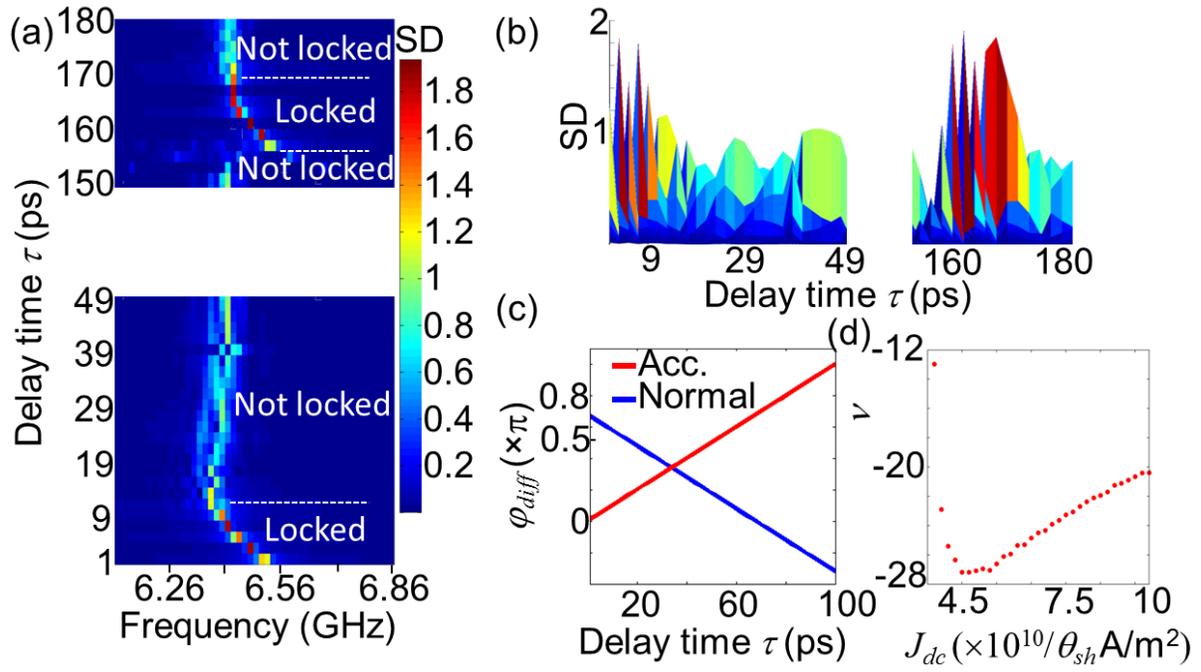

Fig. 3

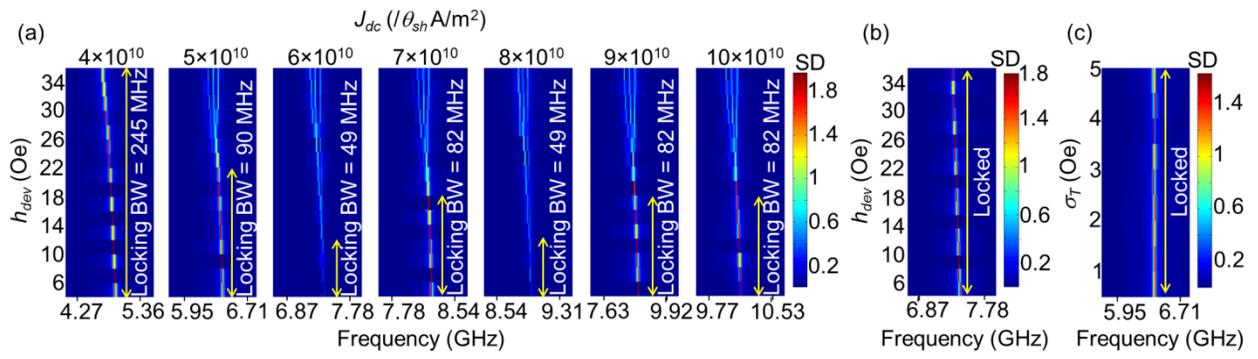

Fig. 4